\documentclass[showpacs,preprintnumbers,aps,prc,floatfix]{revtex4}
\usepackage{longtable}
\usepackage{graphicx}
\usepackage{dcolumn}
\begin{document}
\title
{Level structure of  $^{103}$Ag at high spins}
\author
{S.~Ray$^{1}$, N.S.~Pattabiraman$^{1}$$^{a}$, Krishichayan$^{1}$,
A.~Chakraborty$^{1}$$^{b}$, S.~Mukhopadhyay$^{1}$, S.S.~Ghugre$^{1}$,
S.N.~Chintalapudi$^{1}$, A.K.~Sinha$^{1}$,
U.~Garg$^{2}$, S.~Zhu$^{2,c}$, B.~Kharraja$^{2}$, and D.~Almehed$^{2}$}
\affiliation
{$^{1}$UGC-DAE Consortium for Scientific Research, Kolkata Centre,
Sector III, LB-8, Bidhan Nagar, Kolkata 700 098, India;\\
$^{2}$Physics Department, University of Notre Dame, Notre Dame IN 46556, USA.\\}
\altaffiliation[$^{a}$Present address: ]
                           {{\it Department of Physics, University of York, York YO10 5DD, UK}} \ \\

\altaffiliation[$^{b}$Present address: ]
                           {{\it Department of Physics, Krishnath College, Berhampore 742101, India}} \

\altaffiliation[$^{c}$Present address: ]
                           {{\it Physics Division, Argonne National Laboratory, Argonne IL 60439 USA}} \
\\

\date{\today}
\vspace*{2 cm}

\begin{abstract}
High spin states in $^{103}$Ag were investigated  with the Gammasphere array,
using
the $^{72}$Ge($^{35}$Cl,$2p2n$)$^{103}$Ag reaction at an incident beam
energy of 135 MeV. A $\Delta J$=1
 sequence with predominantly
magnetic transitions and two nearly-degenerate
$\Delta J=1$ doublet bands have been observed.
The dipole band shows a decreasing trend in the $B(M1)$
strength as function of spin, a well established feature of magnetic bands.
The nearly-degenerate band structures satisfy the three experimental signatures
  of
chirality in the
nuclei; however microscopic calculations are 
indicative of a magnetic phenomenon.

\end{abstract}
\maketitle
\section{Introduction}
                                                                                
Nuclei in the $A \sim 105$ region with $Z \sim 50$ and $N \ge 56$ have
 low-lying states containing one or more proton holes in the high-$\Omega$
$g_{9/2}$ orbitals, and neutrons in the low-$\Omega$ $g_{7/2},d_{5/2}$
and $h_{11/2}$ orbitals. These nuclei are also known to be $\gamma$-soft
 (triaxial), with small deformations
 ($\beta_{2} \sim 0.15$) ~\cite{prad2,gadea,jenkins,vaman}. 
 These conditions are favorable for the observation of (a) magnetic or
 ``shears'' band,
characterized by a $\Delta$ J=1 sequence, with
strong intra-band (M1) transitions, weak crossover (E2) transitions,
and decreasing $B(M1)$ strengths with increasing
rotational frequency; and,
(b) chiral  bands, exhibiting nearly degenerate
energy states, with a small, but almost constant, energy difference 
between the bands.
A spontaneous breaking of the chiral symmetry can take place for 
configurations where the angular momenta of the valence protons,
the valence neutrons, and the core are mutually perpendicular.
This can occur, for example, when the proton and neutron Fermi 
levels are located in the lower part of valence proton high-j 
(particle like) and in the upper part of valence neutron 
high-j (hole like) subshells, and the core is triaxial.

 Recently, P. Dutta {\it et al.} \cite{prad1} have reported
 the possibility of
co-existence of the principal and tilted axis rotation in $^{103}$Ag.
The possible configuration suggested for these bands was,
$\pi [(g_{9/2})]\otimes \nu [(g_{7/2})(h_{11/2})]$ \cite{prad1}.
 Also, in the neighboring
nucleus $^{104}$Ag, strong intra-band (M1) transitions have been observed with weak 
cross over (E2) transitions \cite{prad2}.
These bands were suggested to be based on the 
$\pi [(g_{9/2})] \otimes \nu [(g_{7/2})(d_{5/2})(h_{11/2})^{2}]$
configuration; the deformation parameters
$\beta_{2}$ and $\gamma$ were found to be $0.17$ and $29^\circ$, respectively.
Similar configuration involving the degenerate $(g_{7/2})/(d_{5/2})$
 orbitals could give rise to ``shears band'' in $^{103}$Ag as well.

The first observation of chiral structures in nuclei were reported 
in $N=75$ isotones in the $A\sim130$ region. Recently 
evidence has been presented for a new region of chirality
 around $A{\sim104}$~\cite{vaman}.
 These chiral partners are based on the ${\pi}(g_{9/2})^{(-1)}
{\otimes}{\nu}{(h_{11/2})}$ configuration. Such configurations are 
energetically favorable in $^{103}Ag$, hence it is 
 of interest to explore the occurrence of chiral bands (structures) 
in this nucleus.

 Magnetic rotation is expected in weakly deformed nuclei, when
high $j_\pi$ particle of one kind combine with high $j_\nu$
 holes of the other kind
 of nucleons. 
Then the vectors $\bf  j_\pi$, $\bf j_\nu$ subtends a large angle
with $\bf J$ (total angular momentum),
 specifying the orientation and combining to a large
 dipole moment. To observe regular bands with many transitions
the vectors $\bf j_\pi$ and $\bf j_\nu$ of the particles and holes should 
gradually align
with $\bf J$. This situation occurs when at least two high $j$ particles
 and one high -$ j$ hole form the rotating dipole. In a shell model
 study of magnetic rotation band, it has been found that regular bands
 only appear if the core of the nucleus is soft enough such that high -$\it j$
particles may induce a slight deformation.
The gradual alignment of 
$\bf j_\pi$, $\bf j_\nu $ with  $\bf J$
 is reminiscent to the closing of the blades
 of a pair of shears, leading to the name ``shears mechanism''.

 This
behavior arises naturally in the framework of the 
tilted axis cranking (TAC) model~\cite{frau1,frau2,frau3,dimi1,dimi2},
 and has also been
 described in an intuitive semi-classical approach~\cite{macchi,macchi1}.
 In the 
latter it is found that the interaction between $\bf j_\pi$ and 
$\bf j_\nu $ can be described as an effective $\bf P_{2}$ force.
The closing of the particle and hole spin vector leads to a 
specific large drop of B(M1) values with increasing spin,
 since $\bf {\overrightarrow {| \mu_{\bot}|}}$ is decreasing
 with a decreasing opening angle 
and B(M1) $\propto$ $ \bf{\overrightarrow{| \mu_{\bot}|^{2}}}$.

The tilted axis cranking (TAC) framework also
 predicts that in moderately deformed nuclei, 
 the angular momentum is generated from the gradual alignment of
spins of quasiparticles (quasineutron or quasiproton) along the direction of
total angular momentum. This leads to
  the violation of signature symmetry
 due to non-Principal axis rotation which gives rise to 
 regular increase in the M1 transition energies, resulting in tilted
 axis rotation.
 We report here on the observation of
 magnetic rotation, coupled with tilted rotation for
 three bands in the nucleus $^{103}$Ag.

\section{Experiment}
                                                            
High-spin states in $^{103}$Ag were populated using  the
$^{72}$Ge($^{35}$Cl,$2p2n$)$^{103}$Ag reaction at an incident beam energy of
 135 MeV using the ATLAS facility at
 Argonne National Laboratory .The target was 1 mg cm$^{-2}$ thick,
evaporated onto a 15 mg cm$^{-2}$-thick gold foil. A thin
(40 $\mu$g cm$^{-2}$) Al layer was evaporated between the target
 and the backing to avoid migration of the target material into the gold.
De-exciting $\gamma$ rays were detected with the
 Gammasphere detector array in its ``stand alone'' mode;
the array comprised of 101 Compton-suppressed Ge detectors at the time of the
experiment.
Events were recorded when at least three suppressed Ge detectors
 in the array detected $\gamma$ rays within the prompt coincidence time window.
 The total coincidence data set
consisted of approximately 2$\times 10^{9}$ ``triple coincidence'' events.
Data were sorted into the conventional (E$_{\gamma}$)$^{N}$
(N=3,4) histograms and were analyzed using
RADWARE \cite{radford} and IUCSORT \cite{nsp1,nsp2,nsp3} software.
\section{Results}
\subsection{Level Scheme}

\begin{figure}
\includegraphics[height=16cm,angle=270]{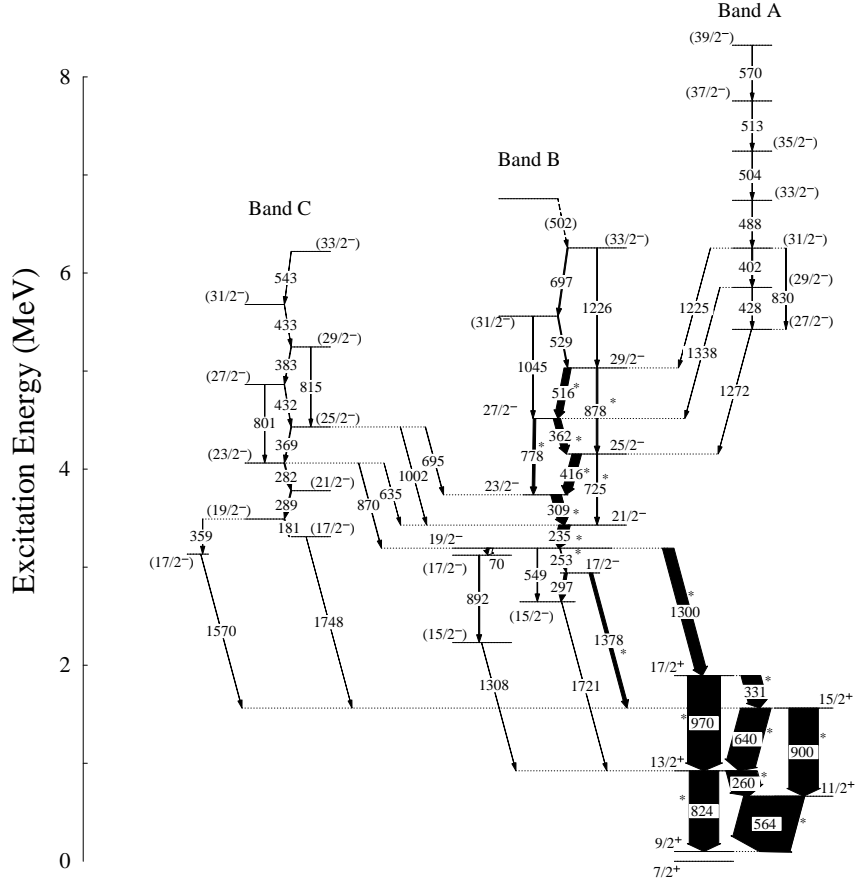}
\caption{\label{fig1}
Partial level scheme for $^{103}$Ag for the levels obtained from the
$^{72}$Ge($^{35}$Cl,$2p2n$)$^{103}Ag$ reaction. Previously-known
transitions are marked with a {\bf (*) }.  The widths of the arrows
are roughly proportional to the corresponding intensities. The spin and parity
 assignments
given in parentheses are tentative. }
\end{figure}

The partial level scheme for $^{103}Ag$ obtained from this experiment is shown in Fig.~\ref{fig1}.
Multipolarity of the de-exciting $\gamma$-rays of $^{103}Ag$ were deduced from 
the $\gamma$-ray coincidences and angular
 correlations\cite{stephans,beausang}.
 The higher fold coincidence data were unfolded into doubles
 and were sorted into the conventional angle-dependent matrices. Two
 dimensional histograms were generated for events where one of the 
$\gamma$- transition was detected in the $90^{\circ}$ ring while it's 
coincidence partner was detected at $31.7^{\circ}$ or $37.3^{\circ}$.

The data from these two rings were summed up for better statistics,
 and the average angle of $35^{\circ}$ 
is used in the subsequent text.
Another matrix was similarly created with coincidence
 data between the $90^{\circ}$ and $145^{\circ}$ detectors
($145^{\circ}$ is the average angle for the detectors
 at $142.6^{\circ}$ or $148.3^{\circ}$ ring).

In our procedure, we define the intensity asymmetry ratio $R_{int}$ as

$$
R_{int} = \frac{ I_ {{\gamma}_1}~ at ~  35^{\circ} and ~145^{\circ};
~ gated~  with~ {\gamma}_2~ at~  90^{\circ}}
{I_ {{\gamma}_1}~ at ~ 90^{\circ}; ~ gated~ with ~{\gamma}_2
~ at ~ 35^{\circ} and ~145^{\circ}}
$$

\begin{figure}
\includegraphics[height=21cm]{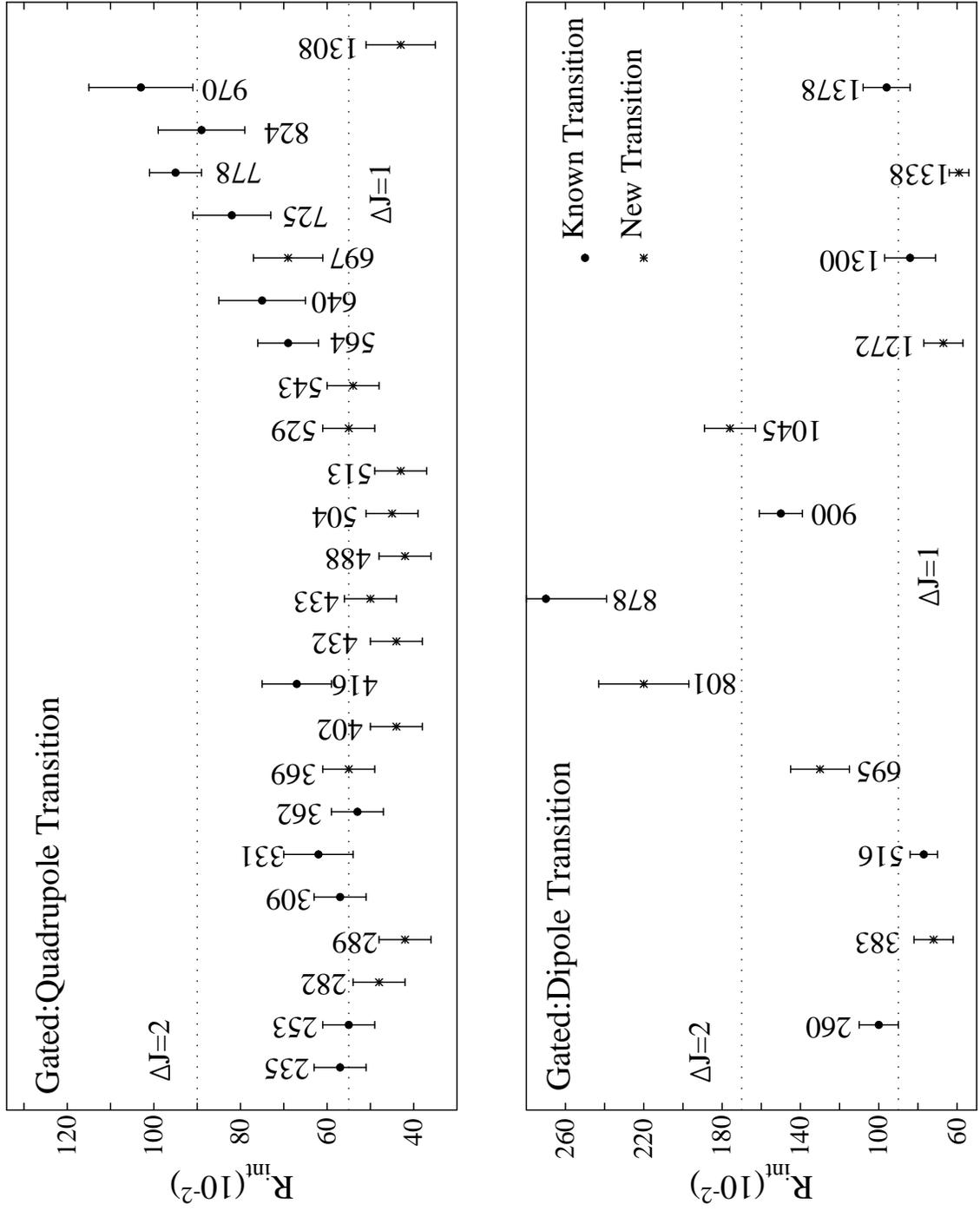}
\caption{\label{fig2}
$R_{int}$ plotted for $\gamma$-ray transitions of $^{103}$Ag, using
 quadrupole transition and dipole transition as gates.
The lines correspond to the value  of $R_{int}$ for the known quadrupoles and
dipoles.
}
\end{figure}

Assuming stretched transitions, the $R_{int}$ value is $\sim$ 0.9
 for quadrupoles and $\sim$ 0.55 for dipoles when gated on an $E(2)$ 
transition. When the gating transition was a dipole the corresponding 
values were $R_{int}$ $\sim$ 1.7 for quadrupole and $\sim$ 0.9 for dipoles.
 These ratios can be used to make reliable spin assignments by 
comparing the present results with those of 
$\gamma$ rays belonging to $^{103}Ag$ and its neighboring nuclei 
whose multipolarity has been established previously.
The presence of cross-over transitions (when observed) corroborated 
these assignments. Representative values of $R_{int}$ are listed in 
Table I and plotted as a function of $\gamma$-ray energy in Fig.~\ref{fig2}.

\begin{figure}
\includegraphics[height=12cm,width=12cm,angle=270]{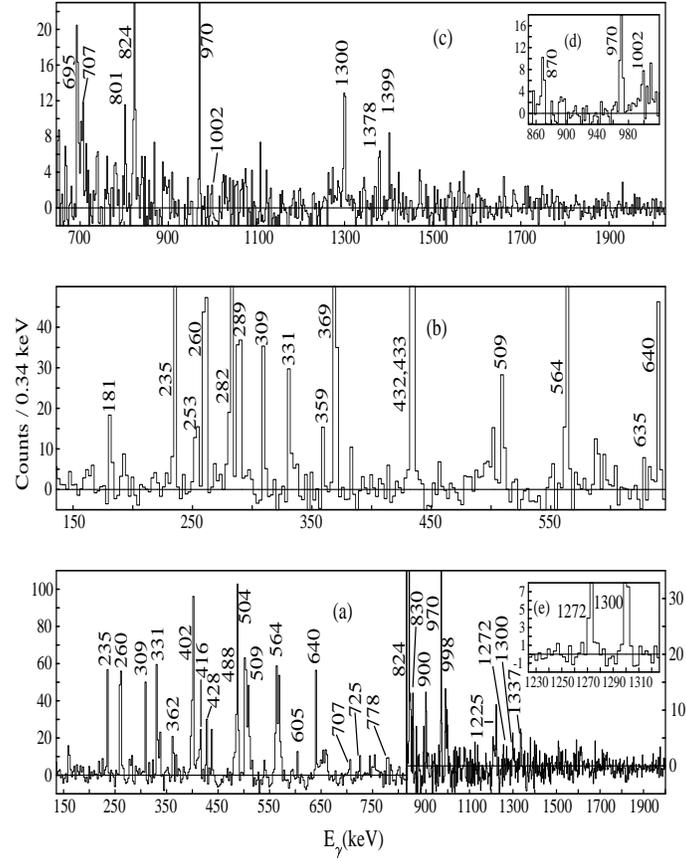}
\caption{\label{fig3}
Background subtracted $\gamma$-$\gamma$-$\gamma$ spectra in $^{103}$Ag
with double gates set on the 383- and 543-keV transitions of
Band C(top and middle panel), and the 570- and 513-keV transitions
 of Band A(lower panel).
The spectrum in inset (d) has been obtained from triple gates on 433-, 383- and 824-keV,
transitions
and the inset (e) from triple gates on the
402-, 428- and 235-keV transitions.
The ``linking'' transitions for both bands are also indicated in the triple-gated spectra.}
\end{figure}

Overall, the level scheme is in good agreement with
the most recent work \cite{prad1} but has been considerably
extended, with the addition of more than 60 new transitions. 
The level sequences of interest are labeled as Band A, Band B and Band C;
Fig.~\ref{fig3} shows the gated coincidence spectra
for Bands C and A.

Band A has been observed for the first time
and consists of a sequence of $\Delta$J=1 transitions of energy(multipolarity)
 428(M1), 402(M1), 488(M1), 504(M1), 513(M1), 570(M1), 830(E2), respectively
(all energies are listed in keV in this paper).
The 1225(M1), 1272(M1) and 1338(M1) transitions connect this band with Band B
which has been extended with the observation of
the following new transitions:
502, 529(M1), 697(M1), 1045(E2) and 1226(E2).

The 1225-, 1338- and the 1272-keV transitions connect the newly
 observed Band A with Band B. Their M1 assignment may be argued as follows:
Let us assume 
all these connecting transitions are E1's.
For an E1 1225-keV transition the multipolarities of the
transitions which de-excite from the 6166.7-keV level would be 402(M1) and 830(E2).  This
would result in a substantial intensity difference (typically about 5 orders of magnitude)
between the 830(E2) and the 1225-keV(E1) transition. However, experimentally
they are found to have nearly the same intensities. Thus, the 1225-keV
 transition
cannot be assigned an E1 multipolarity.

If the 1338-keV transition were an E1, this would imply an E1 also
 for the 402-keV ($\Delta J=1$) transition and an M2 multipolarity for
the cross-over 830-keV transition, which is not favored by lifetime arguments.
Similar arguments hold true for the 1272-keV transition as well. 

When gated by a sum gate of 235-and 309-keV
($\Delta J=1$) M1 transitions, we obtained an $R_{int}$ value of 0.67 for
 the 1272-keV transition.
These gates are free from any contamination from neighboring nuclei .
 Thus the
1272-keV transition involves a $\Delta J=1$ change in angular momentum. 
The average $R_{int}$ value of known M1 transitions using the same gate was $\sim 0.7$.
Similar values were obtained for established M1 transitions in the neighboring nuclei.
Further, using the same gate a value of $\sim 0.9$ was obtained for established
 E1 transitions. Similar values have been obtained for E1 
transitions in the neighboring $^{104}Ag$ nucleus, also.
 Hence the 1272-keV transition is tentatively
assigned as an M1 transition and the band-head for
 the Band B is assigned
$J^\pi =27/2^{-}$.

We have observed a weak crossover transition of 830-keV
(whose multipolarity could not be determined) for the lowest two members
of the band.
However, the 402-and 428-keV transitions have $\Delta J=1$; these are assigned M1 multipolarity.
The $J^\pi$=$29/2^{-}$ assignment is further corroborated by the observation of
an M1 transition of 1338 keV ($29/2^{-}_{A} \rightarrow$$27/2^{-}_{B}$
where the subscript denotes the corresponding band).
The present statistics did not permit us to assign the multipolarity for the 570-keV
 transition; all in-band transitions have been assigned M1 multipolarity, however,
 based on the results for the more intense transitions in the lower part of the band.
                                                                                                      
Band B was known up to $J^{\pi} =29/2^{-}$ and $E_{x}= 5470.7$ keV \cite{prad1}.
The present investigation
has identified five new members belonging to this band: 502(M1), 529(M1), 697(M1),
 1045(E2) and 1226-keV(E2).
The angular correlation procedure employed in this work
 successfully reproduced the multipolarity of the previously-reported
 transitions.
The coincidence relations indicated that 1226- and 1045-keV were crossover transitions
and the corresponding $\Delta J=1$ transitions are of 516, 529 and 697 keV.
The $R_{int}$ values indicated that the 529-keV and 697-keV transitions are indeed
 $\Delta J=1$. We could not deduce the multipolarity for the
1045- and 1226-keV transitions due to the presence of $\gamma$-rays of similar energy, 
in yrast and other
non-yrast bands of $^{103}Ag$. However, the occurrence of $M1$ transitions 
 (516 keV and 529 keV) helped us assign an $E2$
nature to the 1045 keV crossover transition and,
on similar basis, the 1226-keV [529-keV (M1) +
697-keV (M1)] transition has also been assigned an $E2$ multipolarity.

The sequence labeled as Band C has been observed for the first time.
The in-band members of this sequence are 181(M1), 282(M1), 289(M1), 359(M1),
 369(M1),  383(M1), 432(M1), 433(M1) 543(M1), 1570(E1) and 1748(E1).
 We have also observed two crossover transitions of
801(E2) and 815 keV(E2).
Band C is linked to Band B with
 635(M1), 695(M1), 870(E2) and 1002(E2) transitions, the
multipolarities of which establish the negative parity of this band.
The present statistics permitted us to assign $\Delta J=1$ for the
695-keV transition. Further, both the 695- and 1002-keV transitions originate
from the same level and connect to levels differing in spin by one unit.
The possibilities for the electromagnetic character for these two transitions
are, then, M1 and E2 or E1 and M2, respectively. The E1 and M2 combination is not likely
 because, these transitions would then have lifetimes that are different
by almost six orders in magnitude. This is  inconsistent
with the experimental observations as both the transitions de-excite
from the same level, and are observed in prompt coincidence
with their respective coincidence $\gamma$-transitions.
 Hence, the transitions have been assigned M1 and E2 multipolarities, respectively, and
the de-exciting level
is tentatively assigned as $J^\pi$=($25/2^{-}$).
A similar situation prevails for the 635- and 870-keV transitions, so
 the corresponding de-exciting level is
tentatively assigned $J^\pi$=($23/2^{-}$).

 The present statistics did not permit us to undertake the
multipolarity assignments for weak 1570- and 1748-keV inter-band transitions.

\subsection{DSAM Analysis}

As mentioned above, Bands A, B and C, have very intriguing properties
 : (a) Band-A exhibits characteristics of magnetic rotation, a
phenomena reported in the neighboring nuclei; and
(b) Band-B and Band-C ($\Delta$J=1, rotational band doublets) show
 experimental fingerprints of chiral-partners.  These are
discussed in detail in the subsequent sections.
 To further explore the nature of the bands,
 we have performed DSAM analysis of the various transitions in these bands.

The target thickness (1 mg $cm^{-2}$ thick, evaporated onto a 15
mg $cm^{-2}$- thick gold foil) provided sufficient stopping 
power to slow down and stop the recoiling nuclei
 and to allow DSAM lifetime measurements
to be performed. Angle-dependent matrices were formed such that 
 one of the matrices had coincidence events from detectors at 
 $31.7^{\circ}$, $37.4^{\circ}$ (average $\sim 35^{\circ}$)
 versus detectors at $90^{\circ}$, and another had events from
 detectors at $142.6^{\circ}$, $148.3^{\circ}$ (average $\sim 145^{\circ}$) 
versus detectors at $90^{\circ}$.

Background subtracted spectra were generated by gating on the 
$90^{\circ}$ detectors. Transitions that exhibited 
Doppler broadened line shapes were fitted using the 
Lineshape code of Wells and Johnson \cite{wells} from which 
lifetimes and the corresponding transition strengths have been 
obtained.

The Lineshape analysis code was used to generate 5000 
Monte Carlo simulations for the velocity history of recoiling 
nuclei traversing through the target and backing material 
in time steps of 0.001 ps. Electronic stopping powers were 
taken from the shell-corrected tabulations of Northcliff 
and Shilling \cite{north}.

The major source of systematic errors in DSAM lifetime measurements are related to
\begin{enumerate}
\item
The uncertainties due to the stopping power parameterization adopted
 to describe the
slowing down process of the recoiling ions.
\item
The prescription followed for the feeding of the band levels from
unobserved transitions {\em viz.} the side feeding.
\end{enumerate}
Other uncertainties would include systematic errors due to the parameterization
of the background, and instrumental errors, if any. However the contribution
from these is not expected to be significant.
                                                                                               
In the present work information on the slowing down was carried out
 using the prescription of Northcliff and Shilling 
 and were repeated using the electronic stopping
powers of Zeigler \cite{zeigler}.
The lifetime values so obtained are
similar, within error limits.
 This value provided us with the typical uncertainties due
 to parameterization
of stopping power.
                                                                                               
The present data has been analyzed using top-gates
 when possible which eliminates
the uncertainties from side-feeding. These values have been used to cross-check
the value when gated from below, thus minimizing the systematic errors
 due to side-feeding.

The uncertainties in lifetimes from the fits
 were determined by a
 statistical method using the subroutine MINOS\cite{james} by varying
 the parameters until the $\chi^{2}$
reaches the lowest value, $\chi^{2}_{min}$; this point determines the
 best-fit parameter values.
The region over which $\chi^{2}$ takes on a value
between $\chi^{2}_{min}$ and
  $\chi^{2}_{min}$+1
corresponds to ``one standard deviation'' or $68\%$ confidence interval.
The uncertainty for a given parameter was found by varying that parameter
in steps above (below) its best value. At each step, this parameter
 was fixed and
$\chi^{2}$ was re-minimized by varying all other parameters. The step at which the
re-minimized $\chi^{2}$ equaled $\chi^{2}_{min}$+1 was used for positive
 (negative)
 uncertainty for this parameter.

\begin{figure}
\includegraphics[height=21cm]{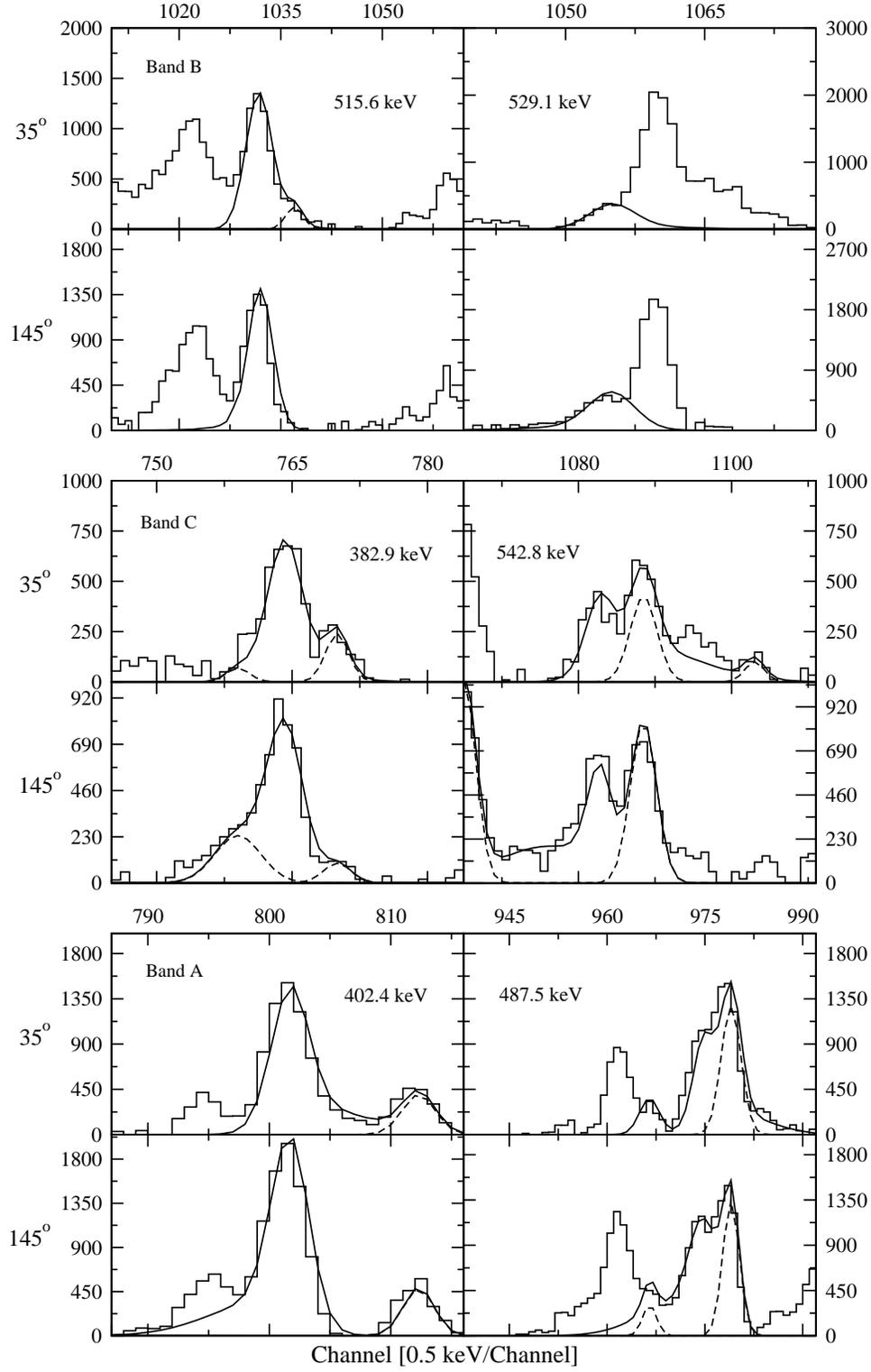}
\caption{\label{fig4}
Experimental and theoretical lineshapes for 383, 402, 488, 516, 529,
and 543 keV $\gamma$- rays at the forward (35$^\circ$; left panels) and backward
(145$^\circ$; right panels) angles with respect to the beam direction. Contaminant peaks 
are shown by dotted lines and theoretical line shape in solid lines.}
\end{figure}

Doppler-broadened lineshapes were clearly observed for all
levels above $J^\pi$=$27/2^{-}$ in Band B and Band C, and for
 the levels above
$J^\pi$=$29/2^{-}$ in Band A. Representative line-shape
fits are depicted in Fig.~\ref{fig4}.

For each band, B(M1) and B(E2) transition rates have been
 extracted using the equations given in \cite{chiara}.
The measured lifetimes along with the corresponding B(M1)
 and B(E2) values are enumerated in Table II where the values correspond
 to the calculations following the
Northcliff and Shilling prescription.  In calculating the
 B(M1) values, a mixing ratio $\delta$=0 was assumed for all
  $\Delta J$=1 transitions.

\section{Discussion}

As pointed out earlier, Band A resembles a
magnetic band, where as Bands B and C exhibit the 
experimental 
fingerprints of
chiral-partners. A number of nearly-degenerate $\Delta J$ = 1 bands have been
observed in the past few years in the
$A \sim 130$ (see, for example, Refs.~\cite{szhu,rainovski,starosta,starosta1})
 and
$A \sim 100$ ~\cite{vaman,joshi,joshi1,timar,joshi2} regions, and have been
proposed as chiral partners.
The main fingerprints of chirality have been identified \cite {vaman} as:

(a) Energy degeneracy between doublet bands;

(b) a constant $S(I)$ parameter, defined as $S(I) = [E(I)-E(I-1)]/2I$,
as a function of spin; and,

(c) staggering of $B(M1)/B(E2)$ values. 

\begin{figure}
\includegraphics[height=21cm,angle=270]{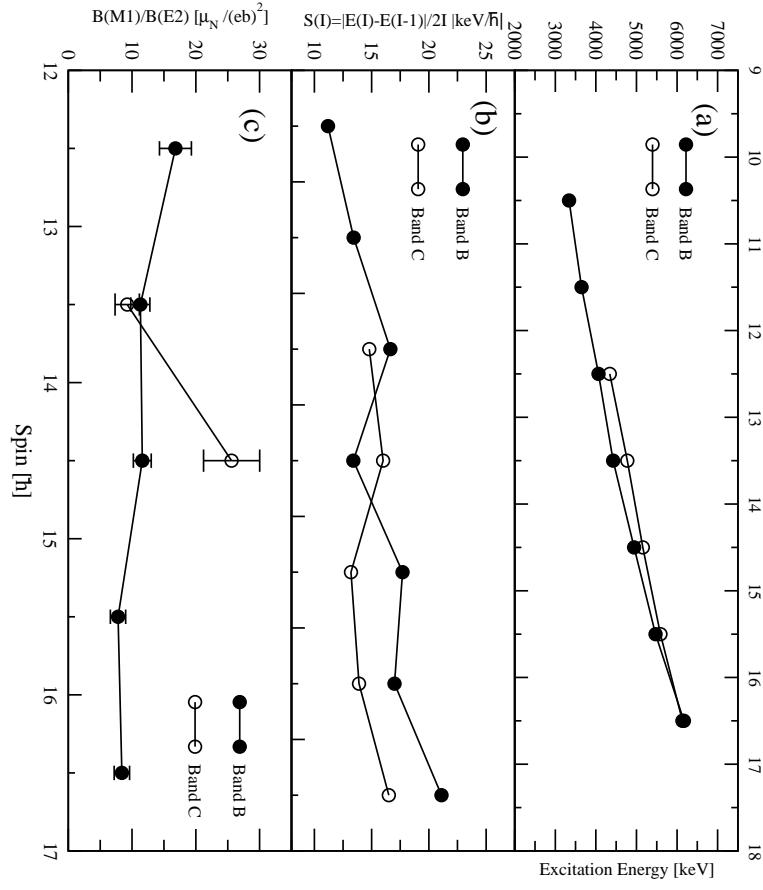}
\caption{\label{fig5}
Chiral fingerprints for Bands B and C in $^{103}Ag$: (a) excitation energy vs spin; 
(b) $S(I)$ vs spin;
(c) $B(M1)/B(E2)$. }
\end{figure}

In Fig.~\ref{fig5}, we plot these three
fingerprints for Bands B and C, and they appear to satisfy all of the
aforementioned criteria.
For example, in Fig.~\ref{fig5}(a), the excitation energies of the bands are
shown as
a function of spin. Where it can be seen that, the energy separation
 between states of the
same spin decreases gradually with increasing spin, eventually becoming nearly
degenerate ($\Delta$E = 19 keV) at spin $33/2^{-}$. Such a behavior is
characteristic of the ``chiral''
bands observed so far and has been attributed to a gradual transition from planar
tilted rotation to aplanar tilted rotation.
 At spins above $J=31/2 \hbar$, it is expected that both
the bands remain close to degenerate in energy.

The second characteristic, which essentially serves as a consistency check,
is the independence of  $S(I)= [E(I)- E(I-1)]/2I$ as a function
of spin $I$. As seen from  Fig.~\ref{fig5}(b),  $S(I)$ is generally independent of
the spin $I$.
In Fig.~\ref{fig5}(c), the $B(M1)/B(E2)$ ratios indicate a certain staggering as a
function of spin.
These observations are very similar to those reported for $^{104}$Rh
\cite{vaman},
and $^{100}$Tc \cite{joshi}, where the corresponding bands have been identified as
chiral doublets.

Traditionally, magnetic and chiral bands have been well understood within the 
 frame work of TAC. In addition, the semi-classical approach of Machiavelli {\it et al.}
 ~\cite{macchi,macchi1}
  provides a qualitative understanding of the phenomenon of magnetic rotation. 
We have applied both these approaches in order to understand the observed
level structure.

Self-consistent cranking model calculations were performed for comparison
 with Band A, B
and C.
Three-dimensional tilted axis cranking (3D TAC) calculations were, then, carried out
 using the same set of parameters
 to investigate the possible existence of magnetic and
 chiral rotation in $^{103}Ag$.
For the three $\Delta J$=1 bands, we found $\theta$$<$ $90^\circ$, which
is in accordance with their $\Delta$ J=1 character.
The deformation parameters, $\epsilon$ and $\gamma$, and the tilt angle
$\theta$ and $\phi$ were obtained for a given frequency $\hbar\omega$ and for
a particular
 configuration, then searching for a local minimum on the multi parameter
surface of the total routhian. All the calculations done below find a minimum
with $\phi=0$ which corresponds to planar TAC solution.
The proton and neutron Fermi surfaces lie too far from the $N=50$, $Z=50$ closed
 shell to warrant using a quasiparticle treatment of the
 protons and neutrons.
 The gap parameter
$\Delta$ was chosen as $80\%$ of the odd-even mass difference $\Delta_{oe}$
for protons and neutrons. The proton and neutron chemical potentials, 
$\lambda_\pi$ and $\lambda_\nu$,
were chosen such that the particle number of $Z\sim47$ and $N\sim56$
 could be reproduced.

\subsection{Band A}                                                                                
Band A exhibits  characteristics of magnetic rotation, a
phenomenon already well established in the neighboring nuclei \cite {prad2}.
As mentioned in Ref. \cite{macchi}, the shears mechanism
is generated by the residual interaction between the proton
and the neutron blades with a strength proportional to
$P_{2}(cos\theta)$, where $\theta$ is the
shears angle.

Experimentally, the shears angle for a pure shears band can be estimated
 at each observed spin $J$, by calculating 
$\theta = cos^{-1}[(J^{2}-j_{\pi}^{2}-j_{\nu}^{2})/2j_{\pi}j_{\nu}]$.
 The excitation energy for each state for pure shears should 
result solely from the closing of blades.

\begin{figure}
\includegraphics[height=18cm,width=12cm,angle=270]{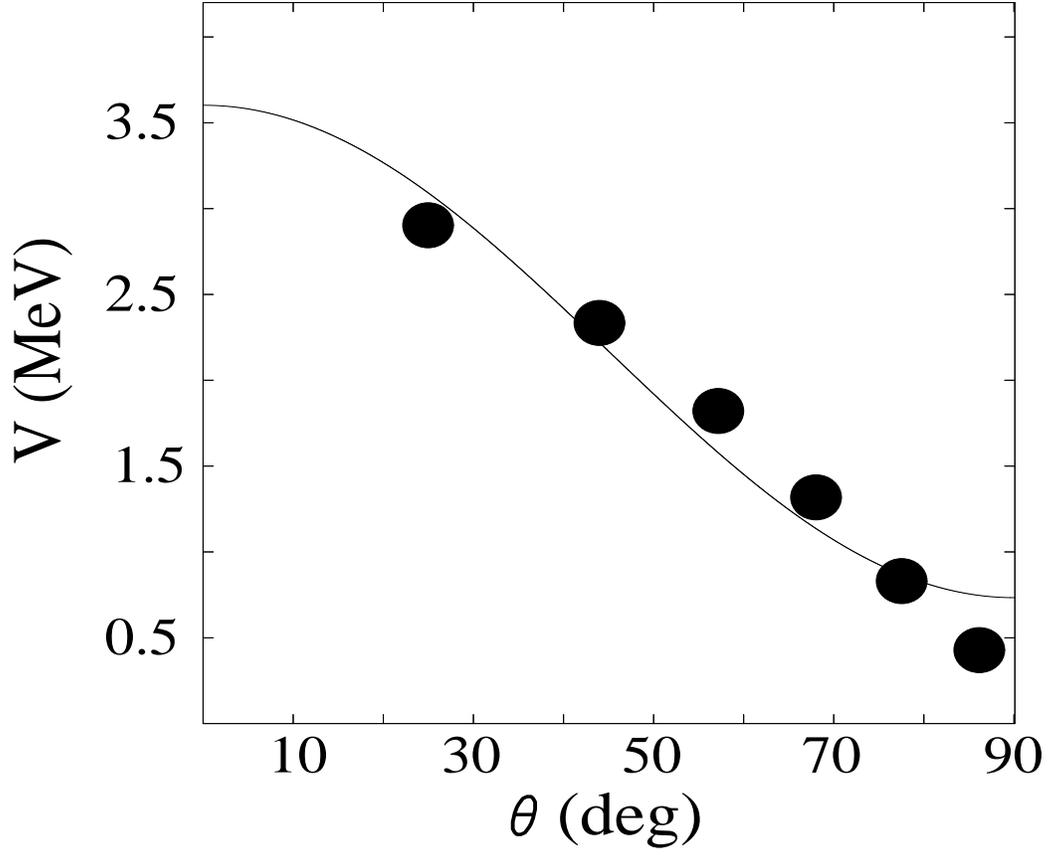}
\caption{\label{fig6}
The effective
 interaction between
angular momentum
vectors $j_{\pi}$ and $j_{\nu}$, as a function of shears angle
$\theta$. The solid curve is the expected dependence of a pure $P_{2}$
term in the interaction, for Band A. }

\end{figure}

 Assuming a pure shears band (with no contribution
from the core), the energies of the states
in this band relative to the $J^\pi =27/2^{-}$ level
were calculated by using the $\pi (g_{9/2}) \otimes \nu (h_{11/2})(g_{7/2})$
configuration.
The highest experimentally observed spin for this sequence is $J = 39/2$.
The energies, $V(\theta)$ are plotted as a function of $\theta$ in
 the form $V(\theta)$=$\frac{1}{2}$$V_{2}(3cos^{2}\theta-1)+V_{0}$,
which is appropriate for a force generated by the exchange
 of a quadrupole phonon
  in Fig.~\ref{fig6}. 
From these results, the strength of the
 interaction was obtained to be
$318\pm40$ keV, which is typical for 
such bands in this region.
 
In Fig.~\ref{fig7}, the aligned angular momentum $i_{x}$ of Band A
 has been plotted as a function of rotational frequency $\omega$.
 The reference parameters for $i_{x}$ were:
$\Im_{0}$=7.0$\hbar^{2}$/MeV and $\Im_{1}$=15.0$\hbar^{4}$/$MeV^{3}$
\cite{regan}.
The aligned angular momentum plot  shows an 
up bend of 4$\hbar$ which continues with increasing $\hbar\omega$.

The TAC calculation for Band A were performed assuming the
 $\pi (g_{9/2})^3$ $\otimes$ $\nu g_{7/2} h_{11/2}$
 five quasiparticle configuration.
The equilibrium value for the deformation parameter is found to be
$\epsilon \sim$ 0.10 with a decreasing  trend along the band.
 The value of $\gamma$ is found to be $19^{\circ}$ and 
 increases along the band.
 The observed spin, energy, and B(M1) values of Band A are
plotted in Fig.~\ref{fig8} together with the results from the TAC
calculation~\cite{frau3,dimi2}. The calculated E(I) plot is shown by the
dashed line whereas the experimental data is represented
 as a solid line. There is a reasonable
 agreement between the experimental values and
TAC calculation for both the energy and the transition rates.
 The experimental I($\omega$) plot (Fig.~\ref{fig8})
 show signs of single particle
 alignment that is not reproduced in
the calculation. The fact that the $\Delta J=2$, E2 transitions are very weak in
the experiment can also be understood from the calculations where the B(E2)
rates are predicted to be close to zero.

The $\Im_2$/B(E2) ratio for Band A is
$\sim$450 $\hbar$$^{2} MeV^{-1}$$(eb)^{-2}$, which is comparable to that of a 
magnetic rotor. This again indicates
 that Band A in $^{103}Ag$ arises due to magnetic rotation. 
The vital fingerprint of the shears mechanism is the characteristic drop in the
magnitude of
$B(M1)$ with increasing rotational frequency because of the rapid closure of
shears.
The experimental $B(M1)$'s are presented in Table II.
The typical values of $B(M1)$ are $\sim$ 3--5 $\mu^{2}_{N}$,
with a decreasing tendency with spin, a signature of magnetic-rotational bands
\cite{frauen}.

\subsection{Bands B and C}

Although Band B and C show
 experimental fingerprints of chiral-partners,
  the tilted axis cranking calculations 
for these bands 
yielded a planar solution ($\phi=0$), essentially ruling out
  chiral
 behavior.                                                                               
 Both these bands
are governed by $g_{9/2}$
proton orbital and $g_{7/2}/d_{5/2}$, $h_{11/2}$ neutron orbitals
 where as Band C has
a different $g_{7/2}/ d_{5/2}$ neutron configuration relative to Band B.
The small deformation ($\epsilon_{2} \sim$0.13) leads to planar tilted
solution with $\gamma$ $\sim$ $15^{\circ}$. Figs.~\ref{fig9},\ref{fig10} show
 that the experimental
energies and the spins of the bands agree with the calculated values.
 While the
transition rates, B(M1), of Band B are very well reproduced by the calculation,
the large B(M1) seen in Band C can not be fully understood. The calculated
B(E2)'s of Band B and C are considerably larger than those in 
 Band A, which would explain why there are substantial E2 transitions in Bands B and C
but not in Band A. Bands B and C show some of the characteristics of a magnetic
band and also have some of the characteristics of a tilted rotational band.

As mentioned earlier, it is tempting to assign Bands B and C as chiral
partners, based on the experimental signatures. However, the
 tilted axis cranking calculation does not support this interpretation.
Instead, the calculation indicates that the two bands are build on
 different quasi-particle
configurations.
The similar characteristics exhibited by these two bands may be due to
 high level density at the neutron Fermi surface, resulting in
several configurations with similar properties. Combined with the high degree 
of symmetry breaking in the mean field solution, this can lead to mixing of 
different quasiparticle configurations which will be manifested in the experimental
 results as an irregular band structure, as for Bands B and C.

\begin{figure}
\includegraphics[height=21cm]{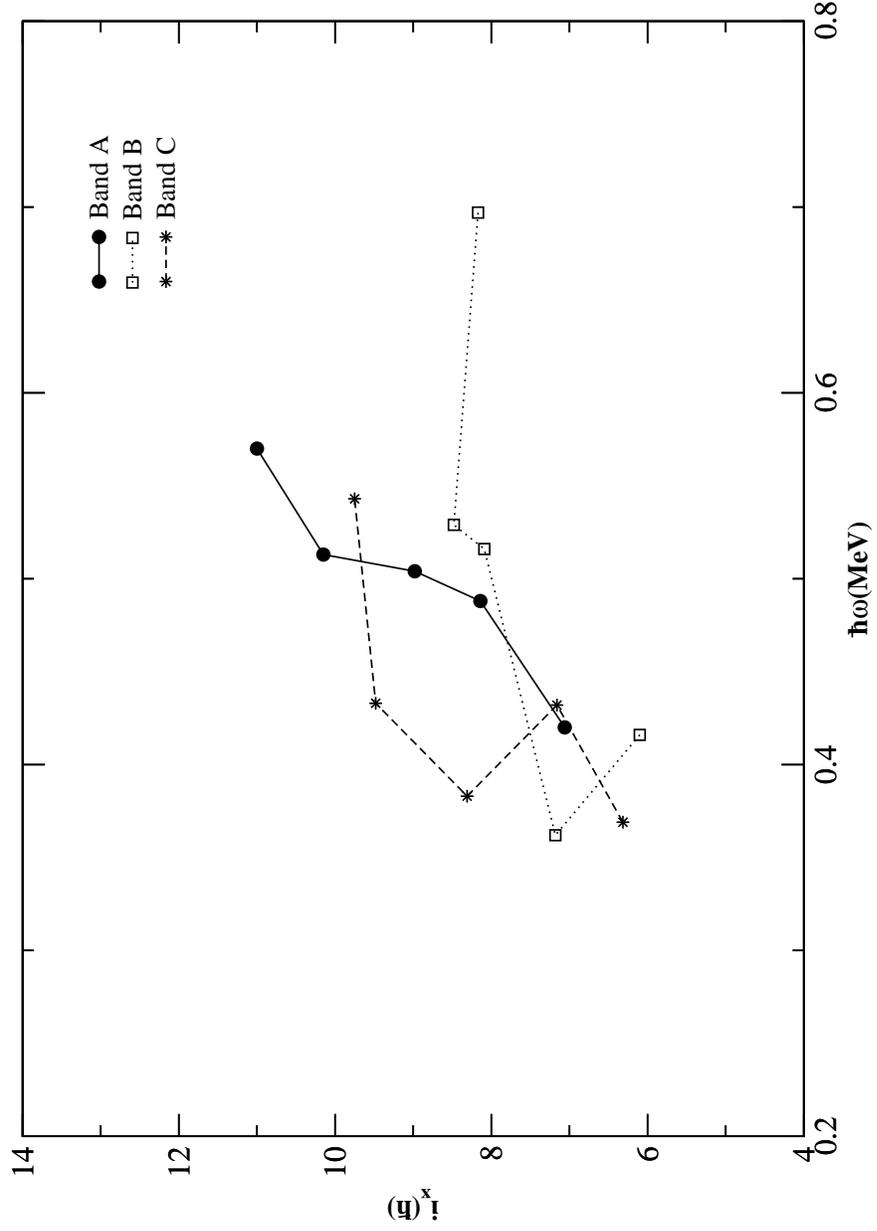}
\caption{\label{fig7}
Aligned angular momentum as a function of rotational frequency
for Bands A, B, C. The experimental frequency is extracted from
the measured $\gamma$ energies using the relation $\hbar\omega$(I)
=$E_{\gamma}$=$E(I)-E(I-1)$. The reference parameters used are:
$\Im_{0}$=7.0$\hbar^{2}$/MeV and $\Im_{1}$=15.0$\hbar^{4}$$/MeV^{3}$}
\end{figure}

\begin{figure}
\includegraphics[height=21cm]{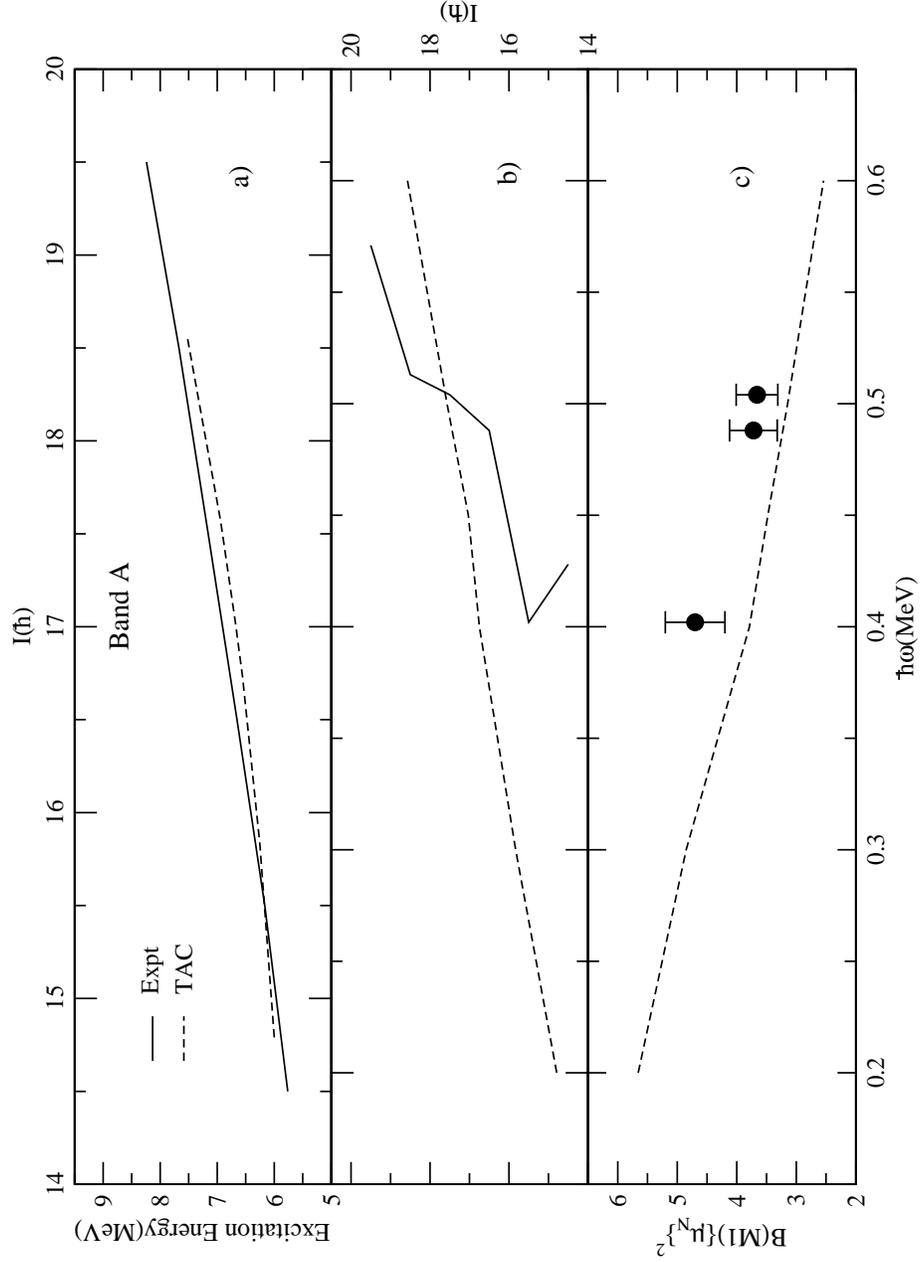}
\caption{\label{fig8}
(a) Excitation energy as a function of angular momentum;
(b) Angular momentum and (c)B(M1) transition rates as a
function of rotational frequency for Band A. The dashed
line in (a), (b) and (c) are results of
TAC calculation for the assigned configuration of Band A.}
\end{figure}

\begin{figure}
\includegraphics[height=21cm]{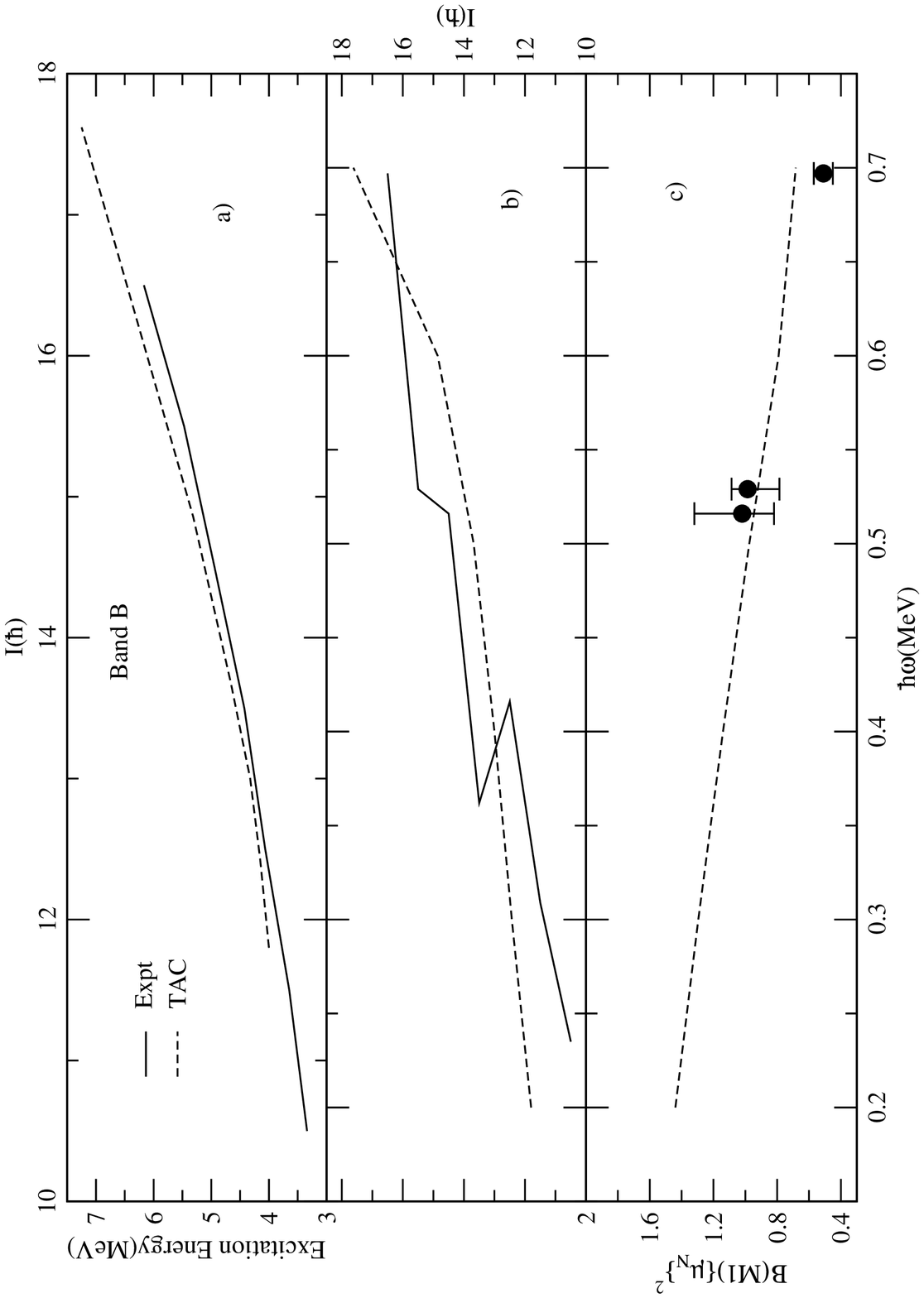}
\caption{\label{fig9}
Same as Fig. 8, but for Band B.}
\end{figure}

\begin{figure}
\includegraphics[height=21cm]{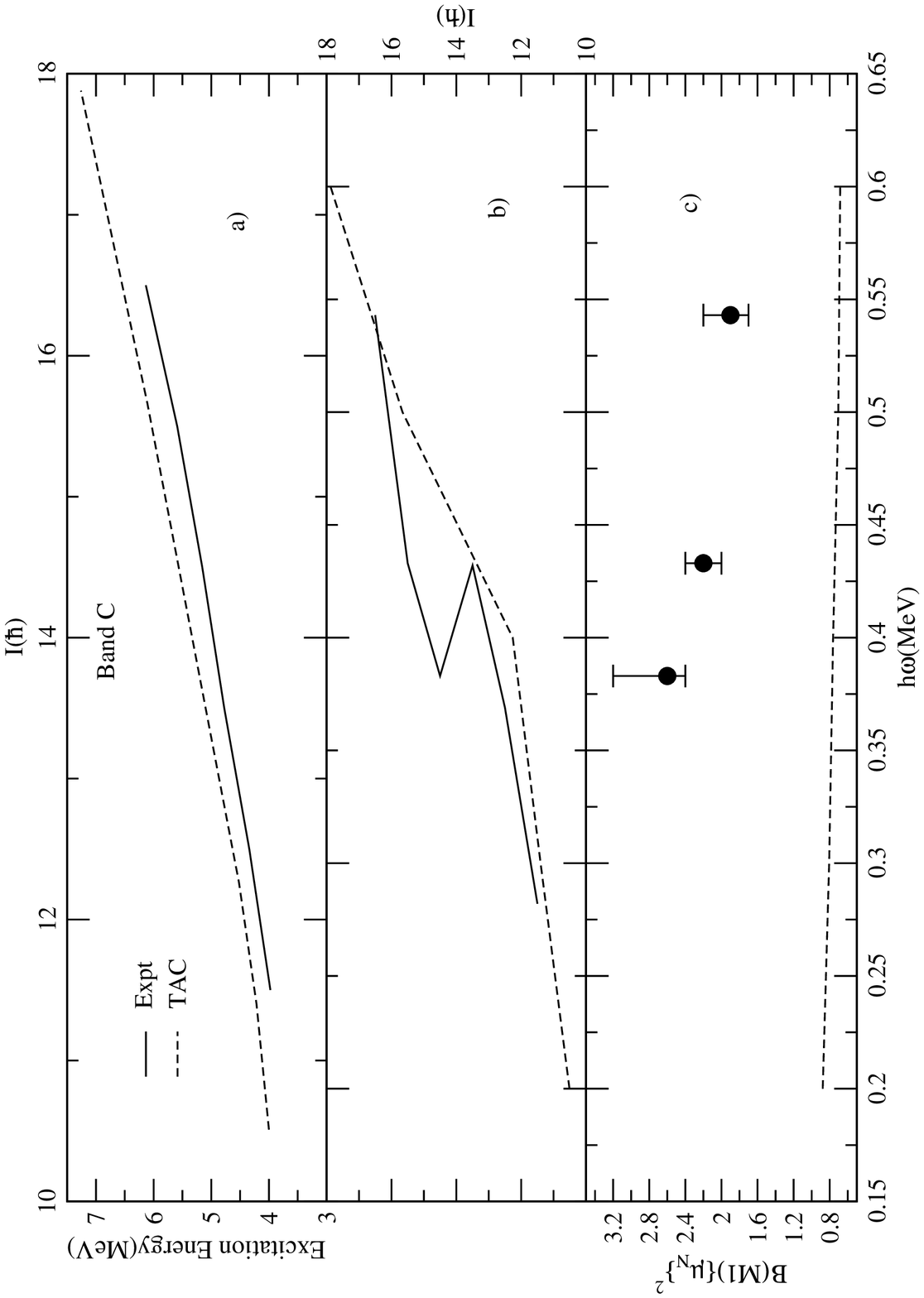}
\caption{\label{fig10}
Same as Fig. 8, but for Band C.}
\end{figure}

For Band B, the aligned angular momentum shows an initial jump of 2$\hbar$
(see Fig.~\ref{fig7})
after which it remains constant at 8 $\hbar$. For Band C, at a rotational 
frequency of approximately 0.35 MeV/$\hbar$ (as shown in Fig.~\ref{fig7})
 the aligned angular momentum
 curve shows a rapid decease in rotational frequency and an increase 
of approximately 3$\hbar$ in aligned angular momentum. This effect can be caused
 by mixing of different configurations or by a partial alignment of
the  $ [(g_{7/2})/(d_{5/2})] $ neutron.

 The $\Im_2$/B(E2) ratio for Bands B and C is 
$\sim$250 $\hbar$$^{2} MeV^{-1}$$(eb)^{-2}$. 
The absence of the aplanar solution and the
 large $\Im_{2}$/B(E2) are supportive of a magnetic rotation for 
the Bands B and C, even though they qualitatively
resemble chiral partners. The smaller value of $\Im_{2}$/B(E2)
and larger deformation of Band B and C relative to Band A indicate that these bands
have a larger component of collective rotation relative to the pure magnetic rotation
of Band A. As pointed out earlier this is also supported by the presence of several $E2$ transitions in 
Band B and C.
The semi classical
 formalism does not work for these bands, also pointing to the presence of
 phenomena other than magnetic rotation.

With the high level density at the neutron Fermi
surface in $^{103}$Ag,
there might be bands with different configurations manifesting very similar
properties in accordance with the chiral fingerprints. Further investigations, both
in theory and experiment, to explore these issues would clearly be useful.                                      

\section{Conclusions}
The level structure of $^{103}$Ag has been substantially extended up to
$E_{x}$= 8240 keV  and $J^{\pi}$=$39/2^{-}$
 with the observation of more than 60 new transitions.
We have observed an intriguing structure in this nucleus,
 {\em viz.} two bands that qualitatively exhibit all the experimental
fingerprints of chiral partners, but do not correspond to an aplanar
 tilted axis cranking solution. Also these bands reveal a larger collective
rotational component than that in the well-established pure magnetic 
rotational bands.
The $\Delta J$=1 sequence (Band A) resembles a
magnetic rotational band. The lifetimes of the states in this band
qualitatively follow the established trend for shears bands.
An interaction strength of $318\pm(40)$ keV has been obtained using the
semi-classical formalism of Macchiavelli {\em et al.}~\cite{macchi,macchi1}.

\section{ACKNOWLEDGMENTS}
We gratefully acknowledge illuminating discussions with Prof. S. Frauendorf and
 Dr. R. V. F. Janssens.
We also thank Prof. J. C. Wells and Mr. P. Dutta for their help with
the computer code LINESHAPE and the ATLAS scientific staff for
 help with the Gammasphere experiment. This work has been supported in part by the
INDO-US, DST-NSF grant (DST-NSF/RPO-017/98), by the U.S. National
Science Foundation under grants no. INT-01115336 and PHY04-57120, and the University 
of Notre Dame.

\newpage
\begin{longtable}{|c|c|c|c|c|c|}
\caption {Gamma transition energy ($E_{\gamma}$) in keV,
Excitation energy ($E_{x}$) in keV,
initial and final spins for the transition,
$R_{int}$, and multipolarity for the states in
 $^{103}$Ag.}
\endfirsthead
\hline
\caption[]{continued...} \\
$E_{\gamma}$ &  $ E_{x} $ & $ J^{\pi}_i   \rightarrow  J^{\pi}_f $& $I_{\gamma}
\footnotemark[1] \footnotemark[2]$&
 $R_{int} $ & Multipolarity\\
\hline
\endhead
\hline
\multicolumn{4}{|r|} {continued...~}\\
\endfoot
\hline
\endlastfoot
\hline
 $E_{\gamma}$ &  $ E_{x} $ & $ J^{\pi}_i   \rightarrow  J^{\pi}_f $ & $I_{\gamma} 
\footnotemark[1] \footnotemark[2]$&
 $R_{int}\footnotemark[3] \footnotemark[4] $ & Multipolarity\\
\hline
 69.9 & 3104.0 & $ 19/2^{-} \rightarrow $ $ (17/2^{-})$ &10.0(10)&  & (M1) \\
 180.8 & 3402.2 & $ (19/2^{-}) \rightarrow $ $  (17/2^{-}) $ &3.4(3)&  &(M1)  \\
 234.9 & 3338.9 & $  21/2^{-} \rightarrow $ $  19/2^{-} $ &36.8(40) &$0.57(6)^q$& M1  \\
 252.5 & 3104.0 & $ 19/2^{-} \rightarrow $ $  17/2^{-} $ &2.8(3)& $0.55(6)^q$ &M1 \\
 260.0 & 833.7 & $ 13/2^{+}\rightarrow $ $ 11/2^{+}$  &100(10)& $1.0(1)^d$ & M1 \\
 282.3 & 3973.9 & $(23/2^{-})\rightarrow $ $(21/2^{-})$ &3.4(3)&  $0.48(6)^q$ & (M1) \\
 289.4 & 3691.0 & $(21/2^{-})\rightarrow $ $ (19/2^{-}) $ &3.8(4) & $0.42(6)^q$ & (M1) \\
 296.6 & 2851.5 & $ (17/2^{-})\rightarrow $ $(15/2^{-} )$ &10.0(10)&  & (M1)  \\
 309.0 & 3647.9 & $ 23/2^{-} \rightarrow $ $ 21/2^{-} $ &34.2(40)& $0.57(6)^q$& M1  \\
 330.5 & 1803.7 & $ 17/2^{+} \rightarrow $ $  15/2^{+} $&63.7(70)& $0.62(8)^q$ &M1 \\
 358.8 & 3402.2 & $ (19/2^{-}) \rightarrow $ $ (17/2^{-}) $ &1.3(1)&  & (M1)  \\
 361.9 & 4426.0 & $27/2^{-}\rightarrow $ $25/2^{-}$ &23.8(20)&  $0.53(6)^q$ &M1   \\
 368.8 & 4340.8 & $ (25/2^{-}) \rightarrow $ $  (23/2^{-})$ &2.3(2)&  $0.55(6)^q$ &(M1)   \\
 382.9 & 5157.0 & $ (29/2^{-})  \rightarrow $ $  (27/2^{-}) $ &1.79(20)&  $0.72(10)^d$ &(M1)\\
 402.4 & 6166.7 & $ (31/2^{-}) \rightarrow $ $ (29/2^{-}) $&3.7(4)& $0.44(6)^q$ &(M1)   \\
 416.2 & 4064.1 & $  25/2^{-} \rightarrow $ $  23/2^{-} $ &31.3(30)& $0.67(8)^q$ & M1\\
 427.8 & 5764.3 & $  (29/2^{-}) \rightarrow $ $ (27/2^{-} )$ &0.50(6)&    & (M1)  \\
 432.2 & 4774.9 & $ (27/2^{-})  \rightarrow  $ $ (25/2^{-}) $ &1.6(1)& $0.44(6)^q$ &(M1)  \\
 433.2 & 5591.0 & $ (31/2^{-})  \rightarrow  $ $ (29/2^{-}) $ &0.87(8)& $0.50(6)^q$ &(M1) \\
 487.5 & 6654.0 & $(33/2^{-})  \rightarrow $    $ (31/2^{-}) $ &1.6(2)& $0.42(6)^q$ &(M1) \\
 502.0 & 6669.7 & $ $$ \rightarrow $ $ (33/2^{-}) $ &$W^{2}$&  & \\
 503.6 &  7157.8 & $ (35/2^{-}) \rightarrow $$  (33/2^{-})   $ &1.5(2)& $0.45(6)^q$ &(M1) \\
 512.6 & 7670.4 & $ (37/2^{-})  \rightarrow $ $  (35/2^{-}) $ &1.4(2)& $0.43(6)^q$ &(M1)  \\
 515.6 & 4941.6 & $  29/2^{-} \rightarrow $ $ 27/2^{-} $ &24.0(20)& $0.77(7)^d$ & M1 \\
 529.1 & 5470.7 &  $ (31/2^{-}) \rightarrow $  $ 29/2^{-}$ &3.42(40)& $0.55(6)^q$ &(M1)\\
 542.8 & 6133.8 & $ (33/2^{-}) \rightarrow $ $ (31/2^{-}) $ &0.70(1)&   $0.54(6)^q$&(M1) \\
 549.1 & 3104.0 & $ 19/2^{-} \rightarrow $ $ (15/2^{-}) $ &2.2(1)& &(E2) \\
 563.7 &  573.7 & $ 11/2^{+} \rightarrow $ $  9/2^{+} $ &200.0& $0.69(7)^q$ &M1 \\
 569.9 & 8240.3 & $ (39/2^{-}) \rightarrow $ $  (37/2^{-})$ &0.30(5)&  &(M1)  \\
 635.0 & 3973.9 & $(23/2^{-})\rightarrow $ $ 21/2^{-} $ &0.18(1)& & (M1)  \\
 639.5 & 1473.2 & $15/2^{+}\rightarrow $ $ 13/2^{+} $ &100(10)& $0.75(10)^q$ &M1   \\
 694.8 & 4342.7 & $(25/2^{-})\rightarrow $ $ 23/2^{-} $ &1.45(20)& $1.30(15)^d$ &(M1) \\
 697.0 & 6167.7 & $(33/2^{-})\rightarrow $ $ (31/2^{-}) $ &4.1(4)& $0.69(8)^q$ &(M1)   \\
 725.2 & 4064.1 & $25/2^{-}\rightarrow $ $ 21/2^{-} $ &3.6(4)& $0.82(9)^q$ &E2  \\
 778.1 & 4426.0 & $27/2^{-}\rightarrow $ $ 23/2^{-} $ &8.7(8)& $0.95(6)^q$ &E2  \\
 801.0 & 4774.9 & $(27/2^{-})\rightarrow $ $ (23/2^{-}) $ &0.50(1)& $2.20(23)^d$& (E2)   \\
 815.1 & 5157.0 & $(29/2^{-})\rightarrow $ $ (25/2^{-}) $ &0.31(1)&  &(E2)  \\
 823.7 & 833.7  & $ 13/2^{+} \rightarrow $ $ 9/2^{+} $ &100(10)& $0.89(10)^q$& E2 \\
 830.2 & 6166.7 & $(31/2^{-})\rightarrow $ $ (27/2^{-}) $ &0.20(2)&  &(E2)  \\
 869.9 & 3973.9 & $(23/2^{-})\rightarrow $ $ 19/2^{-} $ &1.58(20)&  &(E2)  \\
 877.5 & 4941.6 & $(29/2^{-})\rightarrow $ $ 25/2^{-} $ &5.5(5)& $2.70(31)^d$&(E2)  \\
 892.0 & 3034.1 & $(17/2^{-})\rightarrow $ $ (15/2^{-}) $ &3.9(4)&  &(M1)  \\
 899.5 & 1473.2 & $ 15/2^{+}\rightarrow $ $ 11/2^{+} $ &100(10)& $1.50(11)^d$& E2 \\
 970.0 & 1803.7 & $ 17/2^{+}\rightarrow $ $ 13/2^{+} $ &100(10)& $1.03(12)^q$ & E2  \\
 1001.9 & 4340.8 & $(25/2^{-})\rightarrow $ $ 21/2^{-} $ &0.80(1)&  &(E2) \\
 1044.7& 5470.7 & $(31/2^{-})\rightarrow $ $ 27/2^{-} $ &2.5(3)& $1.76(13)^d$ & (E2)\\
 1225.1& 6166.7 & $(31/2^{-})\rightarrow $ $ (29/2^{-}) $ &0.60(6)&   &(M1) \\
 1226.1& 6167.7 & $(33/2^{-})\rightarrow $ $ (29/2^{-}) $ &2.8(3)&    &(E2)\\
 1272.4& 5336.5 & $(27/2^{-})\rightarrow $ $ 25/2^{-} $ &0.80(1)& $0.67(10)^d$&(M1)   \\
 1300.3& 3104.0 & $ 19/2^{-}\rightarrow $ $ 17/2^{+} $ &38.1(30)& $0.84(13)^d$ & E1  \\
 1308.4& 2142.1 & $ (15/2^{-})\rightarrow $ $ 13/2^{+} $ &1.1(1)& $0.43(8)^q$ &(E1)\\
 1338.3& 6166.7 & $ (29/2^{-}) \rightarrow $ $ 27/2^{-} $ &2.7(3)& $0.59(5)^d$ &(M1)   \\
 1378.3& 2851.5 & $17/2^{-} \rightarrow $ $ 15/2^{+} $ &11.9(10) &$0.96(12)^d$ & E1  \\
 1570.2& 3043.4 & $(17/2^{-})\rightarrow $ $ 15/2^{+} $ &1.1(1)& &(E1)   \\
 1721.2& 2554.9 & $(15/2^{-})\rightarrow $ $ 13/2^{+} $ &1.8(2)&  &(E1)  \\
 1748.2& 3221.4 & $(17/2^{-})\rightarrow $ $ (15/2^{+}) $&0.20(2) & &(E1)  \\
 
\footnotetext[1] { The quoted errors on intensities encompass errors
 due to background subtraction, fitting, and efficiency correction.}
\footnotetext[2] { {\it W} indicates weak transitions whose
 intensity could not be computed.}
\footnotetext[3] {$^{q}$ $R_{int}$ obtained from gate on stretched quadrupole transition.}
\footnotetext[4] {$^{d}$ $R_{int}$ obtained from gate on stretched dipole transition.}

\end{longtable}

\newpage
\begin{longtable}{|c|c|c|c|c|c|}
\caption {Measured level lifetimes and the corresponding B(M1)
and B(E2) rates in $^{103}Ag$. The quoted uncertainties include
the fitting errors and errors in side-feeding intensities}
\endfirsthead
\hline
 Band & $E_{x}$ & $ J^{\pi} $ & $\tau $  & B(M1) &
 B(E2) \\
& (keV) & & (ps) & ($\mu_{N}$$^{2}$) & $(eb)^{2}$\\
\hline
\endhead
\hline
\endfoot
\hline
\endlastfoot
\hline
Band & $E_{x}$  &  $\tau $ &$ J^{\pi} $ & B(M1) &
 B(E2) \\
& (keV)  & (ps) & &($\mu_{N}$$^{2}$) & $(eb)^{2}$\\
\hline
 Band B& 4941.6 & $0.401^{+.11}_{-.06}$& $(29/2)^{-}$&$1.02^{+.30}_{-.20}$ & $0.09^{+.02}_{-.03}$ \\
  &5300.9 & $0.385^{+.04}_{-.06}$& $(31/2)^{-}$& $0.986^{+.10}_{-.20}$  & $0.12^{+.01}_{-.01}$\\
 &6167.7 & $0.330^{+.04}_{-.03}$ &$(33/2)^{-}$& $0.51^{+.06}_{-.06}$   & $0.05^{+.005}_{-.005}$ \\
Band C & 5157.0 & $0.390^{+.03}_{-.07}$ &$(29/2)^{-}$& $2.6^{+.60}_{-.20}$  & $0.09^{+.04}_{-.01}$\\
  &5591.0 & $0.32^{+.02}_{-.03}$&$(31/2)^{-}$& $2.2^{+.20}_{+.20}$   &   \\
 & 6133.8 & $0.19^{+.01}_{-.03}$&$(33/2)^{-}$& $1.9^{+.30}_{-.20}$   &   \\
Band A&6166.7 & $0.18^{+.02}_{-.02}$ &$(31/2)^{-}$& $4.70^{+.50}_{+.50}$  &  \\
  & 6654.0 & $0.13^{+.01}_{-.01}$&$(33/2)^{-}$& $3.72^{+.40}_{-.40}$  &  \\
 & 7157.8 & $0.12^{+.01}_{-.01}$ &$(35/2)^{-}$& $3.66^{+.35}_{-.35}$  &  \\
                                                                                                 
\end{longtable}
\newpage


\newpage


\end{document}